\documentclass[osajnl,twocolumn,showpacs,superscriptaddress,10pt]{revtex4-1} 

\usepackage[T1]{fontenc}
\newcommand{\e}{\mathrm{e}}
\newcommand{\ic}{i}
\newcommand{\B}{\boldsymbol}

\newcommand*\conj[1]{
  \vbox{
  \hrule height 0.3pt
  \kern0.5ex
  \hbox{
  \kern-0.4em
  \ifmmode#1\else\ensuremath{#1}\fi
   \kern-0.em
  }
 } 
}

\usepackage{graphicx}
 \usepackage[utf8x]{inputenc}
\usepackage{amssymb,amsmath}
 \usepackage{siunitx}
\usepackage[overload]{empheq}
\usepackage{subfigure}
\usepackage{numprint}
\usepackage{bm}
\usepackage{pgfplots}

 \usepackage{tikz}
  \usetikzlibrary{%
      decorations.pathreplacing,%
      decorations.pathmorphing,%
  shapes%
  }
\PassOptionsToPackage{dvipsnames}{xcolor}
	\RequirePackage{xcolor} 
\definecolor{halfgray}{gray}{0.55} 
\definecolor{webgreen}{rgb}{0,.5,0}
\definecolor{webbrown}{rgb}{.6,0,0}

  \definecolor{lgreen} {RGB}{180,210,100}
  \definecolor{dblue}  {RGB}{20,66,129}
  \definecolor{lred}   {RGB}{220,0,0}
  \definecolor{nred}   {RGB}{224,0,0}
  \definecolor{norange}{RGB}{210,129,34}
  \definecolor{nyellow}{RGB}{255,221,0}
  \definecolor{ngreen} {RGB}{98,158,31}
  \definecolor{dgreen} {RGB}{78,138,21}
  \definecolor{nblue}  {RGB}{28,130,185}
  \definecolor{jblue}  {RGB}{20,50,100}
  \usepackage[active]{srcltx}

\usepackage{lipsum}

\begin{document}

\title{Transmission enhancement through square coaxial apertures arrays in metallic film: when leaky modes filter infrared light}

\author{Benjamin Vial}
\affiliation{Centrale Marseille, Aix Marseille Université, CNRS, Institut Fresnel, UMR 7249, 13013 Marseille, France}
\affiliation{Silios Technologies, ZI Peynier-Rousset, rue Gaston Imbert Prolong\'ee, 13790 Peynier, France}
\email[]{benjamin.vial@fresnel.fr}

\author{Fr\'ed\'eric Bedu}
\author{Hervé Dallaporta}
\affiliation{Aix Marseille Université, CNRS, CiNaM, UMR 7325, Campus de Luminy, Case 913, 13288 Marseille Cedex 9, France}

\author{Mireille Commandré}
\author{Guillaume Dem\'esy}
\author{André Nicolet}
\author{Frédéric Zolla}

\affiliation{Aix Marseille Université, CNRS, Centrale Marseille, Institut Fresnel, UMR 7249, 13013 Marseille, France}

\author{St\'ephane Tisserand}
\author{Laurent Roux}
\affiliation{Silios Technologies, ZI Peynier-Rousset, rue Gaston Imbert Prolong\'ee, 13790 Peynier, France}

\begin{abstract}We consider arrays of square coaxial apertures in a gold layer and study their diffractive 
behavior in the far infrared region. These structures exhibit a resonant transmission enhancement that 
is used to design tunable bandpass filters. We provide a study of their spectral features and show by a modal analysis 
that the resonance peak is due to the excitation of leaky modes of the open photonic structure. 
Fourier transform infrared (FTIR) spectrophotometry transmission measurements of samples 
deposited on Si substrate show good agreement with numerical results and demonstrate angular tolerance 
up to 30 degrees of the fabricated filters.
\end{abstract}

\maketitle 

\begin{figure}[htbp!]
\includegraphics[width=1\columnwidth]{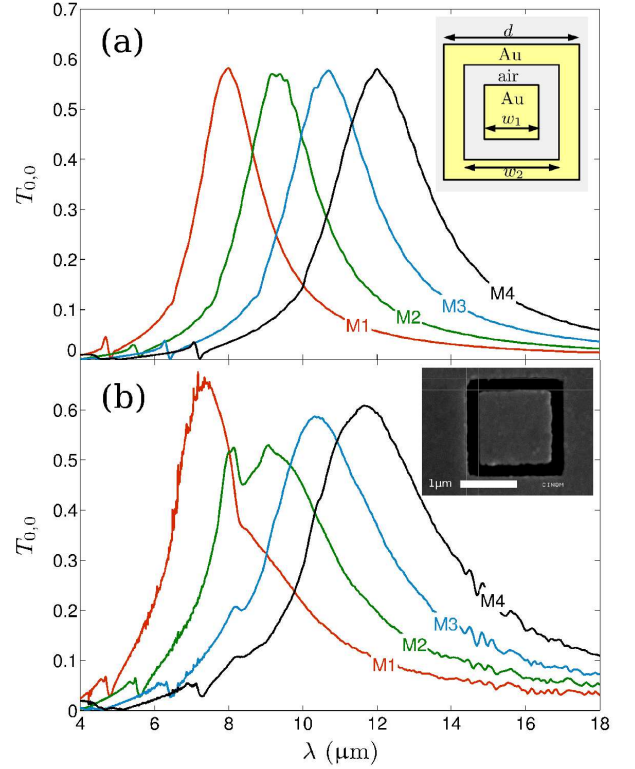}
\caption{Transmission spectra at normal incidence in the specular order $T_{0,0}$ as a function of incident wavelength $\lambda$ 
for the different filters. (a): FEM simulations, (b): FTIR measurements.}
\label{spectresT}
\end{figure}

Nanostructuration 
of metallic surfaces at subwavelength scale can lead to spectacular resonant effects \cite{RevModPhys.82.729}.
In particular, annular aperture arrays (AAAs) have recently drawn considerable attention 
because of their potential application as highly integrated photonic components since 
the work of Baida \textit{et al.} \cite{moreau2003lts,Baida2004,Salvi2005,vanlabeke,Baida2006,poujet}
demonstrating enhanced transmission properties in such structures. AAAs have also been designed and studied experimentally 
at mid-IR wavelengths by Fan \textit{et al.} \cite{Fan2005,Fan2005a}. 
Besides the calculation of transmission spectra, our approach to 
study the resonant phenomena in such metamaterials is to compute the eigenmodes and eigenfrequencies of such 
open electromagnetic systems. This modal approach leads to significant insights into the properties of
metamaterials \cite{Lalanne2006,Grigoriev2013} and eases the conception of diverse optical devices
 \cite{fehrembach,Ding2004a} because it provides a simple picture of 
the resonant processes at stake.\\
In this Letter, we consider arrays of square coaxial apertures in a gold film deposited on a silicon substrate that 
are designed to produce bandpass transmission filters for multispectral imaging applications in the far infrared region. 
The apertures have interior and exterior width denoted $w_1$ and $w_2$ respectively and are arranged in 
a square array of period $d$ (See the inset in Fig.~\ref{spectresT} (a)). The thickness of the metallic film is $h=\SI{90}{\nano\meter}$ which is higher than the 
skin depth of gold in the investigated wavelength range so that the unstructured layer is optically opaque.\\
In a first step, we investigate transmission properties of AAAs by numerical simulations. A Finite Element Method (FEM) 
formulation already described in Refs. \cite{Demesy2009,Demesy2010} is used to solve the so-called \emph{diffraction problem}. 
The array is illuminated from the air superstrate by 
a plane wave of radial angle $\theta_0$, azimuthal angle $\varphi_0$ and polarization angle $\psi_0$. 
The permittivity of gold is described by a Drude-Lorentz model \cite{Ordal1985} and the refractive index of silicon 
is taken from tabulated data \cite{palik}. All materials are assumed to be non magnetic ($\mu_r=1$).
We model a single period with Bloch conditions applied in $x$ and $y$ directions of periodicity such that the electric field 
$\B E$ is quasiperiodic: $\B E(x+d_x,y+d_y,z)=\B E(x,y,z)\e^{\ic(\alpha d_x+\beta d_y)}$, where $\alpha=-k_0\sin \theta_0 \,\cos \varphi_0$ and 
$\beta=-k_0\sin \theta_0 \,\sin \varphi_0$ are the tangential components along $x$ and $y$ of the incident wavevector $\B k_0$. 
Perfectly Matched Layers (PMLs) are used in the $z$ direction normal to the grating in order to damp propagating waves \cite{Berenger1,VialOE2012}. \\
Because of the resonant nature of the transmission spectra, we performed a modal analysis \cite{VialPRA2014} of these structures. 
We use a FEM formulation to solve the \emph{spectral problem}, \textit{i.e.} to find leaky modes
 associated with complex eigenfrequencies $\omega_n=\omega'_n+\ic\omega''_n$ 
of the open nanoresonators \cite{VialMIM}, the real part corresponding to the resonant frequency and the 
imaginary part to the linewidth of the resonance. 
The quasimodes are an intrinsic property of the system and are very useful to characterize the resonant process at stake. 
This approach allows us to compute quickly the features of the resonance and their evolution when the geometric parameters of the 
AAA are changed. The FEM formulation is analog to the diffraction problem except that there are no sources. Bloch conditions are applied 
with fixed real quasiperiodicity coefficients $\alpha$ and $\beta$ along $x$ and $y$ and PMLs are used in the direction orthogonal to the array. 
As a first approximation, the materials are assumed to be non dispersive, which makes the spectral problem linear. 
 To take into account dispersion, the eigenvalue problem is solved iteratively with updated values of permittivity. 
This procedure converges rapidly due to the slow variations of the permittivity of the considered materials in the 
far infrared range.\\
\begin{figure}[htbp!]
\includegraphics[width=1\columnwidth]{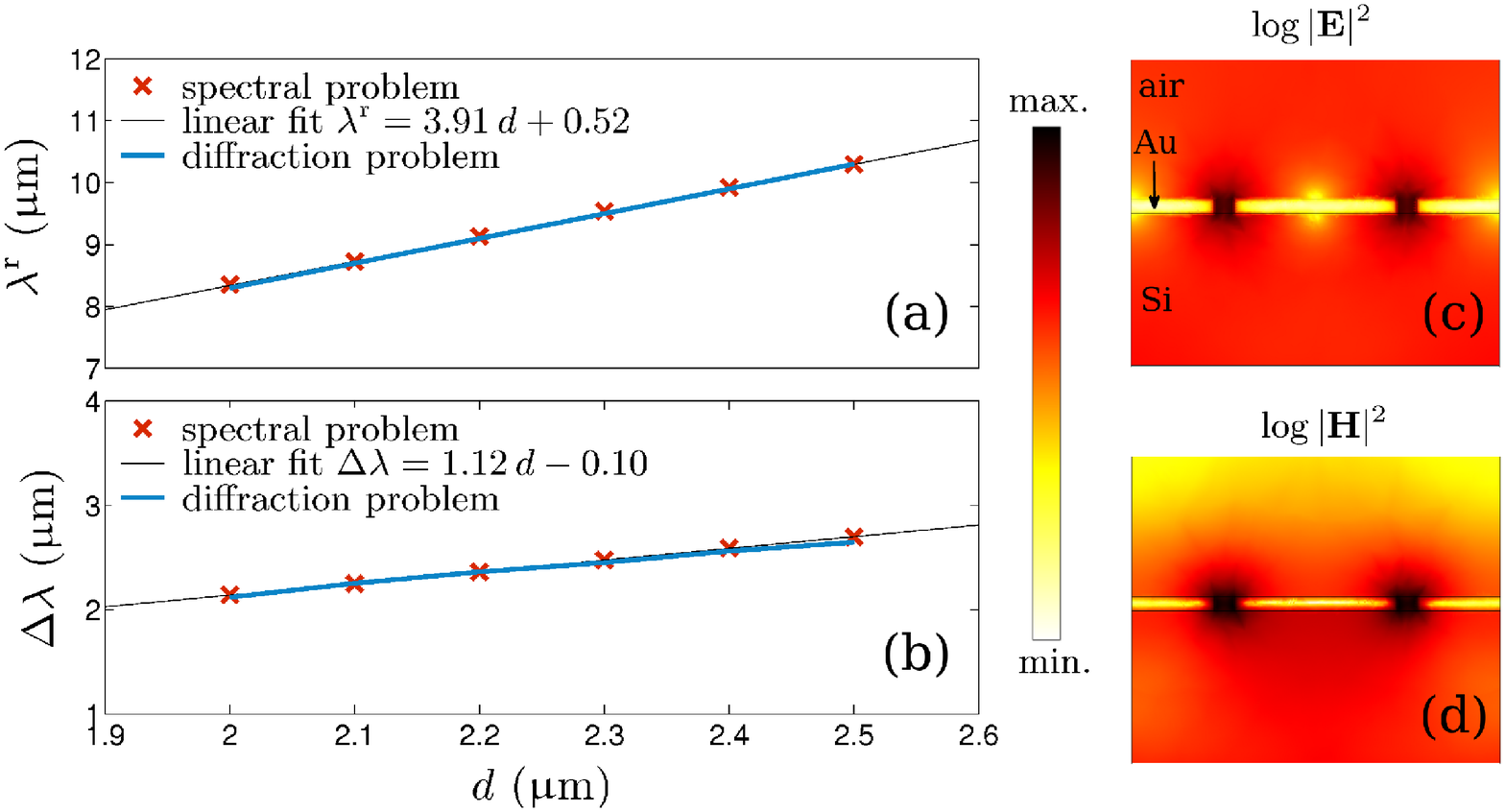}
\caption{Spectral parameters of the resonance as a function of the period $d$ obtained from
calculated transmission spectra (blue line) and extracted from the degenerated 
eigenfrequencies (red crosses). (a): resonant wavelength, (b): spectral width. Electric (c) and magnetic (d) 
field maps of the leaky mode in the $Oyz$ plane for $d=\SI{2.4}{\micro\meter}$.\label{paramspec}}
\end{figure}
We solved the spectral problem with $\alpha=\beta=0$ for structures with 
$f_1=w_1/w_2=0.8$, $f_2=w_2/d=0.55$ and various periods $d$. 
Because of the symmetry of the problem, we find two degenerated eigenmodes 
corresponding to TE and TM polarization associated with the same eigenfrequency $\omega_1$. 
We have represented on Fig.~\ref{paramspec} (a) the evolution of the resonant wavelength 
$\lambda^{\rm r}=2\pi c/\omega'_1$ as a function of $d$ for the TM mode (red crosses). 
The evolution is linear and fits well with the formula $\lambda^{\mathrm r}=3.91\,d+0.52$ (thin black line).
We also extracted the resonant wavelength corresponding to the maximum values of 
calculated transmission spectra, and reported it as a function of $d$ on Fig.~\ref{paramspec} (top) (thick blue line). 
The agreement with the results of the spectral problem is excellent. In addition, the imaginary part of the 
eigenfrequency $\omega_1''$ is equal to $2\Delta\omega$, where $\Delta\omega$ is the spectral width of the 
resonance. We can thus obtain the spectral width 
in terms of wavelength from the expression $\Delta\lambda=-4\pi c \omega''_1/{\omega'_1}^2$ and plot it 
as a function of $d$ on Fig.~\ref{paramspec} (b), red crosses). The evolution is also linear (fitted by 
$\Delta\lambda=1.12\,d-0.10$, thin black line) and 
matches the values of the full width at half maximum obtained from the calculated transmission spectra (thick blue line). Field maps corresponding 
to the electric and magnetic field intensity of this resonant mode for $d=\SI{2.4}{\micro\meter}$ are plotted on Fig.~\ref{paramspec} (c) and (d) respectively. 
The electromagnetic field is concentrated in the annular apertures, which confirms that this mode behaves
 as a waveguide-like mode \cite{Baida2004}. These results demonstrate that the 
resonant transmission enhancement is principally due to the excitation of a single leaky mode supported by the AAA.\\
The modal analysis allows us to design four filters with given central wavelength $\lambda^{\rm r}=8$, $9.7$, $10.3$ 
and $\SI{12}{\micro\meter}$ thanks to the linear fit.
We thus obtain the geometric parameters of the filters denoted M1, M2, M3, M4 respectively: 
$d=1920$, $2250$, $2600$, $\SI{2930}{\nano\meter}$, $w_1=850$, 
$990$, $1140$, $\SI{1290}{\nano\meter}$ and 
$w_2=1060$, $12240$, $1430$, $\SI{1610}{\nano\meter}$. These structures are 
homothetic such that $f_1\simeq0.8$ and $f_2\simeq0.55$. Calculated transmission spectra in the specular 
diffraction order $T_{0,0}$ under normal incidence in TM polarization 
are reported in Fig~\ref{spectresT} (a). Note that the order $(n,m)$ is evanescent for $\lambda>d\sqrt{\varepsilon^-/(n^2+m^2)}$ but 
for all filters, the transmission associated with propagative orders other than $(0,0)$ is very weak (<1.5\%) on the 
studied spectral range. The transmission show a resonant peak whose central wavelength can be redshifted by 
increasing solely the lateral dimensions of the AAA. The spectral width varies from 2 to $\SI{3}{\micro\meter}$ as lateral dimensions increase 
while the maximum transmission value remains unchanged around 0.58.\\
Samples with the aforementioned parameters have been fabricated on the same double size polished (100) pure intrinsic silicon substrate. 
The nanostructuration have been performed by electronic lithography using a lift-off process with a negative tone resist to obtain 
a relatively large nanostructured area of $3\times \SI{3}{\milli\meter^2}$ for each of the four samples 
to allow the measurements of FTIR specta.
 Transmission spectra of the samples have been recorded with a Thermo Fisher-Nicolet 6700 Fourier Transform InfraRed (FTIR) spectrophotometer. 
Measurements were performed with a focused polarized light beam with $\pm \SI{16}{\degree}$ divergence and a 
spot diameter of $\SI{1.3}{\milli\meter}$. A rotating deck allows us to tilt the sample in order to record transmission spectrum 
for incident angles between $0$ and $\SI{90}{\degree}$. All the spectra are normalized with a background recorded 
with the substrate alone. The measured spectra at normal incidence and in TM polarization 
are reported on Fig~\ref{spectresT} (b) and show good agreement with the simulated ones, even 
if the experimental resonances are slightly blueshifted and broader because of variations
 on the fabricated aperture widths. The transmission levels at resonance are of the order of 0.6 which is consistent 
with the numerical simulations. Note that another resonance occurs near $\SI{8.3}{\micro\meter}$ for each transmission spectrum 
which could be attributed to a parasitic large scale pattern due to the fabrication process that has been observed in SEM images 
of the samples.\\
\begin{figure}[htbp!]
\includegraphics[width=1\columnwidth]{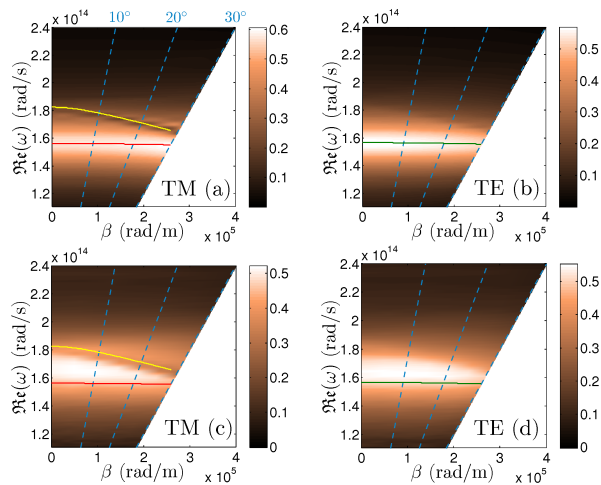}
\caption{Transmission diagrams for the filter M4. Colormap: transmission coefficient 
in the specular order $T_{0,0}$ as a function of frequency $\omega$ and transverse wavenumber $\beta$.
Simulations: (a) TM and (b) TE polarization; measurements: (c) TM and (d) TE polarization. 
The solid lines represent the dispersion relation of the corresponding excited leaky modes.\label{DD}}
\end{figure}
Finally, we study numerically and experimentally the angular behavior of the filters. 
FEM simulations of the diffraction problem for incident angles between $0$ and $30$ degrees are reported 
as a transmission diagram in the space $(\beta,\omega)$ on Fig.~\ref{DD} for TM (a) end TE (b) 
polarization. We also computed eigenfrequencies for different values of $\beta$ and superposed the 
so-called dispersion diagram for the corresponding modes on Fig.~\ref{DD}.
In both cases, the position of the transmission resonance peak is almost constant 
while its intensity slightly increases (resp. decreases) in TM (resp. TE) polarization. These principal resonances 
can be attributed to the excitation of the previously studied waveguide mode because the position of 
the real part of the associated eigenvalue matches the position of the transmission maximum 
(See red curve on Fig.~\ref{DD} (a) for the TM mode and green curve on Fig.~\ref{DD} (b) for the TE mode). 
The spectral width of the transmission peak is also almost constant with 
$\beta$ which is confirmed by the study of the imaginary parts of the eigenfrequencies. 
In addition, a secondary resonant dip appears only in TM polarization and its position redshifts when $\beta$ increases. 
This dip is due to the excitation of the delocalized surface plasmon mode whose dispersion diagram is 
reported on Fig.~\ref{DD} (a) (yellow curve). The displacement of the dip 
with $\beta$ matches the evolution of the real part of the eigenfrequency associated with this mode.\\
Measurements are in good agreement with the simulated ones (See Figs.~\ref{DD} (c) and (d)), even if the experimental principal resonance 
is slightly blueshifted as remarked before. Moreover the transmission dip in TM case is also present on the 
experimental data. Both numerical and experimental results demonstrate that this kind of AAA based filters provide an angular tolerance of 30 degrees.\\
To conclude, we have studied both numerically and experimentally the resonant transmission properties of AAA-based 
transmission bandpass filters in the far infrared. The transmission characteristics of the filter can be tuned by 
changing the value of transverse geometric parameters 
of the structures and the different filters can be fabricated on the same substrate using the same process. 
A modal analysis allows us to infer this transmission peak from the excitation of a leaky mode. The measured spectra are in good agreement with 
FEM simulations, and demonstrate the angular tolerance up to 30 degrees of the fabricated filters.
\begin{acknowledgments}
 This research was supported
by the fundings of PACA regional council and the FEDER program. 
The plasmonic structures were fabricated in the Planete CT-PACA platform.
\end{acknowledgments}


\begin{thebibliography}{21}%
\makeatletter
\providecommand \@ifxundefined [1]{%
 \@ifx{#1\undefined}
}%
\providecommand \@ifnum [1]{%
 \ifnum #1\expandafter \@firstoftwo
 \else \expandafter \@secondoftwo
 \fi
}%
\providecommand \@ifx [1]{%
 \ifx #1\expandafter \@firstoftwo
 \else \expandafter \@secondoftwo
 \fi
}%
\providecommand \natexlab [1]{#1}%
\providecommand \enquote  [1]{``#1''}%
\providecommand \bibnamefont  [1]{#1}%
\providecommand \bibfnamefont [1]{#1}%
\providecommand \citenamefont [1]{#1}%
\providecommand \href@noop [0]{\@secondoftwo}%
\providecommand \href [0]{\begingroup \@sanitize@url \@href}%
\providecommand \@href[1]{\@@startlink{#1}\@@href}%
\providecommand \@@href[1]{\endgroup#1\@@endlink}%
\providecommand \@sanitize@url [0]{\catcode `\\12\catcode `\$12\catcode
  `\&12\catcode `\#12\catcode `\^12\catcode `\_12\catcode `\%12\relax}%
\providecommand \@@startlink[1]{}%
\providecommand \@@endlink[0]{}%
\providecommand \url  [0]{\begingroup\@sanitize@url \@url }%
\providecommand \@url [1]{\endgroup\@href {#1}{\urlprefix }}%
\providecommand \urlprefix  [0]{URL }%
\providecommand \Eprint [0]{\href }%
\providecommand \doibase [0]{http://dx.doi.org/}%
\providecommand \selectlanguage [0]{\@gobble}%
\providecommand \bibinfo  [0]{\@secondoftwo}%
\providecommand \bibfield  [0]{\@secondoftwo}%
\providecommand \translation [1]{[#1]}%
\providecommand \BibitemOpen [0]{}%
\providecommand \bibitemStop [0]{}%
\providecommand \bibitemNoStop [0]{.\EOS\space}%
\providecommand \EOS [0]{\spacefactor3000\relax}%
\providecommand \BibitemShut  [1]{\csname bibitem#1\endcsname}%
\let\auto@bib@innerbib\@empty
\bibitem [{\citenamefont {Garcia-Vidal}\ \emph {et~al.}(2010)\citenamefont
  {Garcia-Vidal}, \citenamefont {Martin-Moreno}, \citenamefont {Ebbesen},\ and\
  \citenamefont {Kuipers}}]{RevModPhys.82.729}%
  \BibitemOpen
  \bibfield  {author} {\bibinfo {author} {\bibfnamefont {F.}~\bibnamefont
  {Garcia-Vidal}}, \bibinfo {author} {\bibfnamefont {L.}~\bibnamefont
  {Martin-Moreno}}, \bibinfo {author} {\bibfnamefont {T.}~\bibnamefont
  {Ebbesen}}, \ and\ \bibinfo {author} {\bibfnamefont {L.}~\bibnamefont
  {Kuipers}},\ }\href {\doibase 10.1103/RevModPhys.82.729} {\bibfield
  {journal} {\bibinfo  {journal} {Rev. Mod. Phys.}\ }\textbf {\bibinfo {volume}
  {82}},\ \bibinfo {pages} {729} (\bibinfo {year} {2010})}\BibitemShut
  {NoStop}%
\bibitem [{\citenamefont {Moreau}\ \emph {et~al.}(2003)\citenamefont {Moreau},
  \citenamefont {Granet}, \citenamefont {Baida},\ and\ \citenamefont {{Van
  Labeke}}}]{moreau2003lts}%
  \BibitemOpen
  \bibfield  {author} {\bibinfo {author} {\bibfnamefont {A.}~\bibnamefont
  {Moreau}}, \bibinfo {author} {\bibfnamefont {G.}~\bibnamefont {Granet}},
  \bibinfo {author} {\bibfnamefont {F.}~\bibnamefont {Baida}}, \ and\ \bibinfo
  {author} {\bibfnamefont {D.}~\bibnamefont {{Van Labeke}}},\ }\href@noop {}
  {\bibfield  {journal} {\bibinfo  {journal} {Opt. Express}\ }\textbf {\bibinfo
  {volume} {11}},\ \bibinfo {pages} {1131} (\bibinfo {year}
  {2003})}\BibitemShut {NoStop}%
\bibitem [{\citenamefont {Baida}\ \emph {et~al.}(2004)\citenamefont {Baida},
  \citenamefont {{Van Labeke}}, \citenamefont {Granet}, \citenamefont
  {Moreau},\ and\ \citenamefont {Belkhir}}]{Baida2004}%
  \BibitemOpen
  \bibfield  {author} {\bibinfo {author} {\bibfnamefont {F.}~\bibnamefont
  {Baida}}, \bibinfo {author} {\bibfnamefont {D.}~\bibnamefont {{Van Labeke}}},
  \bibinfo {author} {\bibfnamefont {G.}~\bibnamefont {Granet}}, \bibinfo
  {author} {\bibfnamefont {A.}~\bibnamefont {Moreau}}, \ and\ \bibinfo {author}
  {\bibfnamefont {A.}~\bibnamefont {Belkhir}},\ }\href {\doibase
  10.1007/s00340-004-1497-3} {\bibfield  {journal} {\bibinfo  {journal} {Appl.
  Phys. B: Lasers Opt.}\ }\textbf {\bibinfo {volume} {79}},\ \bibinfo {pages}
  {1} (\bibinfo {year} {2004})}\BibitemShut {NoStop}%
\bibitem [{\citenamefont {Salvi}\ \emph {et~al.}(2005)\citenamefont {Salvi},
  \citenamefont {Roussey}, \citenamefont {Baida}, \citenamefont {Bernal},
  \citenamefont {Mussot}, \citenamefont {Sylvestre}, \citenamefont {Maillotte},
  \citenamefont {{Van Labeke}}, \citenamefont {Perentes}, \citenamefont {Utke},
  \citenamefont {Sandu}, \citenamefont {Hoffmann},\ and\ \citenamefont
  {Dwir}}]{Salvi2005}%
  \BibitemOpen
  \bibfield  {author} {\bibinfo {author} {\bibfnamefont {J.}~\bibnamefont
  {Salvi}}, \bibinfo {author} {\bibfnamefont {M.}~\bibnamefont {Roussey}},
  \bibinfo {author} {\bibfnamefont {F.~I.}\ \bibnamefont {Baida}}, \bibinfo
  {author} {\bibfnamefont {M.-P.}\ \bibnamefont {Bernal}}, \bibinfo {author}
  {\bibfnamefont {A.}~\bibnamefont {Mussot}}, \bibinfo {author} {\bibfnamefont
  {T.}~\bibnamefont {Sylvestre}}, \bibinfo {author} {\bibfnamefont
  {H.}~\bibnamefont {Maillotte}}, \bibinfo {author} {\bibfnamefont
  {D.}~\bibnamefont {{Van Labeke}}}, \bibinfo {author} {\bibfnamefont
  {A.}~\bibnamefont {Perentes}}, \bibinfo {author} {\bibfnamefont
  {I.}~\bibnamefont {Utke}}, \bibinfo {author} {\bibfnamefont {C.}~\bibnamefont
  {Sandu}}, \bibinfo {author} {\bibfnamefont {P.}~\bibnamefont {Hoffmann}}, \
  and\ \bibinfo {author} {\bibfnamefont {B.}~\bibnamefont {Dwir}},\ }\href
  {\doibase 10.1364/OL.30.001611} {\bibfield  {journal} {\bibinfo  {journal}
  {Opt. Lett.}\ }\textbf {\bibinfo {volume} {30}},\ \bibinfo {pages} {1611}
  (\bibinfo {year} {2005})}\BibitemShut {NoStop}%
\bibitem [{\citenamefont {Labeke}\ \emph {et~al.}(2006)\citenamefont {Labeke},
  \citenamefont {G{\'e}rard}, \citenamefont {Guizal}, \citenamefont {Baida},\
  and\ \citenamefont {Li}}]{vanlabeke}%
  \BibitemOpen
  \bibfield  {author} {\bibinfo {author} {\bibfnamefont {D.~V.}\ \bibnamefont
  {Labeke}}, \bibinfo {author} {\bibfnamefont {D.}~\bibnamefont {G{\'e}rard}},
  \bibinfo {author} {\bibfnamefont {B.}~\bibnamefont {Guizal}}, \bibinfo
  {author} {\bibfnamefont {F.~I.}\ \bibnamefont {Baida}}, \ and\ \bibinfo
  {author} {\bibfnamefont {L.}~\bibnamefont {Li}},\ }\href@noop {} {\bibfield
  {journal} {\bibinfo  {journal} {Opt. Express}\ }\textbf {\bibinfo {volume}
  {14}},\ \bibinfo {pages} {11945} (\bibinfo {year} {2006})}\BibitemShut
  {NoStop}%
\bibitem [{\citenamefont {Baida}\ \emph {et~al.}(2006)\citenamefont {Baida},
  \citenamefont {Belkhir}, \citenamefont {Labeke},\ and\ \citenamefont
  {Lamrous}}]{Baida2006}%
  \BibitemOpen
  \bibfield  {author} {\bibinfo {author} {\bibfnamefont {F.~I.}\ \bibnamefont
  {Baida}}, \bibinfo {author} {\bibfnamefont {A.}~\bibnamefont {Belkhir}},
  \bibinfo {author} {\bibfnamefont {D.~V.}\ \bibnamefont {Labeke}}, \ and\
  \bibinfo {author} {\bibfnamefont {O.}~\bibnamefont {Lamrous}},\ }\href
  {\doibase 10.1103/PhysRevB.74.205419} {\bibfield  {journal} {\bibinfo
  {journal} {Phys. Rev. B}\ }\textbf {\bibinfo {volume} {74}},\ \bibinfo
  {pages} {205419} (\bibinfo {year} {2006})}\BibitemShut {NoStop}%
\bibitem [{\citenamefont {Poujet}\ \emph {et~al.}(2007)\citenamefont {Poujet},
  \citenamefont {Salvi},\ and\ \citenamefont {Baida}}]{poujet}%
  \BibitemOpen
  \bibfield  {author} {\bibinfo {author} {\bibfnamefont {Y.}~\bibnamefont
  {Poujet}}, \bibinfo {author} {\bibfnamefont {J.}~\bibnamefont {Salvi}}, \
  and\ \bibinfo {author} {\bibfnamefont {F.~I.}\ \bibnamefont {Baida}},\
  }\href@noop {} {\bibfield  {journal} {\bibinfo  {journal} {Opt. Lett.}\
  }\textbf {\bibinfo {volume} {32}},\ \bibinfo {pages} {2942} (\bibinfo {year}
  {2007})}\BibitemShut {NoStop}%
\bibitem [{\citenamefont {Fan}\ \emph {et~al.}(2005{\natexlab{a}})\citenamefont
  {Fan}, \citenamefont {Zhang}, \citenamefont {Malloy},\ and\ \citenamefont
  {Brueck}}]{Fan2005}%
  \BibitemOpen
  \bibfield  {author} {\bibinfo {author} {\bibfnamefont {W.}~\bibnamefont
  {Fan}}, \bibinfo {author} {\bibfnamefont {S.}~\bibnamefont {Zhang}}, \bibinfo
  {author} {\bibfnamefont {K.~J.}\ \bibnamefont {Malloy}}, \ and\ \bibinfo
  {author} {\bibfnamefont {S.~R.~J.}\ \bibnamefont {Brueck}},\ }\href {\doibase
  10.1364/OPEX.13.004406} {\bibfield  {journal} {\bibinfo  {journal} {Opt.
  Express}\ }\textbf {\bibinfo {volume} {13}},\ \bibinfo {pages} {4406}
  (\bibinfo {year} {2005}{\natexlab{a}})}\BibitemShut {NoStop}%
\bibitem [{\citenamefont {Fan}\ \emph {et~al.}(2005{\natexlab{b}})\citenamefont
  {Fan}, \citenamefont {Zhang}, \citenamefont {Minhas}, \citenamefont
  {Malloy},\ and\ \citenamefont {Brueck}}]{Fan2005a}%
  \BibitemOpen
  \bibfield  {author} {\bibinfo {author} {\bibfnamefont {W.}~\bibnamefont
  {Fan}}, \bibinfo {author} {\bibfnamefont {S.}~\bibnamefont {Zhang}}, \bibinfo
  {author} {\bibfnamefont {B.}~\bibnamefont {Minhas}}, \bibinfo {author}
  {\bibfnamefont {K.~J.}\ \bibnamefont {Malloy}}, \ and\ \bibinfo {author}
  {\bibfnamefont {S.~R.~J.}\ \bibnamefont {Brueck}},\ }\href {\doibase
  10.1103/PhysRevLett.94.033902} {\bibfield  {journal} {\bibinfo  {journal}
  {Phys. Rev. Lett.}\ }\textbf {\bibinfo {volume} {94}},\ \bibinfo {pages}
  {033902} (\bibinfo {year} {2005}{\natexlab{b}})}\BibitemShut {NoStop}%
\bibitem [{\citenamefont {Lalanne}\ \emph {et~al.}(2006)\citenamefont
  {Lalanne}, \citenamefont {Hugonin},\ and\ \citenamefont
  {Chavel}}]{Lalanne2006}%
  \BibitemOpen
  \bibfield  {author} {\bibinfo {author} {\bibfnamefont {P.}~\bibnamefont
  {Lalanne}}, \bibinfo {author} {\bibfnamefont {J.~P.}\ \bibnamefont
  {Hugonin}}, \ and\ \bibinfo {author} {\bibfnamefont {P.}~\bibnamefont
  {Chavel}},\ }\href {http://jlt.osa.org/abstract.cfm?URI=jlt-24-6-2442}
  {\bibfield  {journal} {\bibinfo  {journal} {J. Lightwave Technol.}\ }\textbf
  {\bibinfo {volume} {24}},\ \bibinfo {pages} {2442} (\bibinfo {year}
  {2006})}\BibitemShut {NoStop}%
\bibitem [{\citenamefont {Grigoriev}\ \emph {et~al.}(2013)\citenamefont
  {Grigoriev}, \citenamefont {Varault}, \citenamefont {Boudarham},
  \citenamefont {Stout}, \citenamefont {Wenger},\ and\ \citenamefont
  {Bonod}}]{Grigoriev2013}%
  \BibitemOpen
  \bibfield  {author} {\bibinfo {author} {\bibfnamefont {V.}~\bibnamefont
  {Grigoriev}}, \bibinfo {author} {\bibfnamefont {S.}~\bibnamefont {Varault}},
  \bibinfo {author} {\bibfnamefont {G.}~\bibnamefont {Boudarham}}, \bibinfo
  {author} {\bibfnamefont {B.}~\bibnamefont {Stout}}, \bibinfo {author}
  {\bibfnamefont {J.}~\bibnamefont {Wenger}}, \ and\ \bibinfo {author}
  {\bibfnamefont {N.}~\bibnamefont {Bonod}},\ }\href {\doibase
  10.1103/PhysRevA.88.063805} {\bibfield  {journal} {\bibinfo  {journal} {Phys.
  Rev. A}\ }\textbf {\bibinfo {volume} {88}},\ \bibinfo {pages} {063805}
  (\bibinfo {year} {2013})}\BibitemShut {NoStop}%
\bibitem [{\citenamefont {Fehrembach}\ and\ \citenamefont
  {Sentenac}(2005)}]{fehrembach}%
  \BibitemOpen
  \bibfield  {author} {\bibinfo {author} {\bibfnamefont {A.~L.}\ \bibnamefont
  {Fehrembach}}\ and\ \bibinfo {author} {\bibfnamefont {A.}~\bibnamefont
  {Sentenac}},\ }\href@noop {} {\bibfield  {journal} {\bibinfo  {journal}
  {Appl. Phys. Lett.}\ }\textbf {\bibinfo {volume} {86}},\ \bibinfo {pages}
  {121105} (\bibinfo {year} {2005})}\BibitemShut {NoStop}%
\bibitem [{\citenamefont {Ding}\ and\ \citenamefont
  {Magnusson}(2004)}]{Ding2004a}%
  \BibitemOpen
  \bibfield  {author} {\bibinfo {author} {\bibfnamefont {Y.}~\bibnamefont
  {Ding}}\ and\ \bibinfo {author} {\bibfnamefont {R.}~\bibnamefont
  {Magnusson}},\ }\href {\doibase 10.1364/OPEX.12.001885} {\bibfield  {journal}
  {\bibinfo  {journal} {Opt. Express}\ }\textbf {\bibinfo {volume} {12}},\
  \bibinfo {pages} {1885} (\bibinfo {year} {2004})}\BibitemShut {NoStop}%
\bibitem [{\citenamefont {Dem\'{e}sy}\ \emph {et~al.}(2009)\citenamefont
  {Dem\'{e}sy}, \citenamefont {Zolla}, \citenamefont {Nicolet},\ and\
  \citenamefont {Commandr\'{e}}}]{Demesy2009}%
  \BibitemOpen
  \bibfield  {author} {\bibinfo {author} {\bibfnamefont {G.}~\bibnamefont
  {Dem\'{e}sy}}, \bibinfo {author} {\bibfnamefont {F.}~\bibnamefont {Zolla}},
  \bibinfo {author} {\bibfnamefont {A.}~\bibnamefont {Nicolet}}, \ and\
  \bibinfo {author} {\bibfnamefont {M.}~\bibnamefont {Commandr\'{e}}},\ }\href
  {\doibase 10.1364/OL.34.002216} {\bibfield  {journal} {\bibinfo  {journal}
  {Opt. Lett.}\ }\textbf {\bibinfo {volume} {34}},\ \bibinfo {pages} {2216}
  (\bibinfo {year} {2009})}\BibitemShut {NoStop}%
\bibitem [{\citenamefont {Dem\'{e}sy}\ \emph {et~al.}(2010)\citenamefont
  {Dem\'{e}sy}, \citenamefont {Zolla}, \citenamefont {Nicolet},\ and\
  \citenamefont {Commandr\'{e}}}]{Demesy2010}%
  \BibitemOpen
  \bibfield  {author} {\bibinfo {author} {\bibfnamefont {G.}~\bibnamefont
  {Dem\'{e}sy}}, \bibinfo {author} {\bibfnamefont {F.}~\bibnamefont {Zolla}},
  \bibinfo {author} {\bibfnamefont {A.}~\bibnamefont {Nicolet}}, \ and\
  \bibinfo {author} {\bibfnamefont {M.}~\bibnamefont {Commandr\'{e}}},\ }\href
  {\doibase 10.1364/JOSAA.27.000878} {\bibfield  {journal} {\bibinfo  {journal}
  {J. Opt. Soc. Am. A}\ }\textbf {\bibinfo {volume} {27}},\ \bibinfo {pages}
  {878} (\bibinfo {year} {2010})}\BibitemShut {NoStop}%
\bibitem [{\citenamefont {Ordal}\ \emph {et~al.}(1985)\citenamefont {Ordal},
  \citenamefont {Bell}, \citenamefont {R.~W.~Alexander}, \citenamefont {Long},\
  and\ \citenamefont {Querry}}]{Ordal1985}%
  \BibitemOpen
  \bibfield  {author} {\bibinfo {author} {\bibfnamefont {M.~A.}\ \bibnamefont
  {Ordal}}, \bibinfo {author} {\bibfnamefont {R.~J.}\ \bibnamefont {Bell}},
  \bibinfo {author} {\bibfnamefont {J.}~\bibnamefont {R.~W.~Alexander}},
  \bibinfo {author} {\bibfnamefont {L.~L.}\ \bibnamefont {Long}}, \ and\
  \bibinfo {author} {\bibfnamefont {M.~R.}\ \bibnamefont {Querry}},\ }\href
  {\doibase 10.1364/AO.24.004493} {\bibfield  {journal} {\bibinfo  {journal}
  {Appl. Opt.}\ }\textbf {\bibinfo {volume} {24}},\ \bibinfo {pages} {4493}
  (\bibinfo {year} {1985})}\BibitemShut {NoStop}%
\bibitem [{\citenamefont {Palik}(1991)}]{palik}%
  \BibitemOpen
  \bibfield  {author} {\bibinfo {author} {\bibfnamefont {E.~D.}\ \bibnamefont
  {Palik}},\ }\href@noop {} {\emph {\bibinfo {title} {{H}andbook of optical
  constants of solids}}}\ (\bibinfo  {publisher} {Academic Press},\ \bibinfo
  {year} {1991})\BibitemShut {NoStop}%
\bibitem [{\citenamefont {Berenger}(1994)}]{Berenger1}%
  \BibitemOpen
  \bibfield  {author} {\bibinfo {author} {\bibfnamefont {J.-P.}\ \bibnamefont
  {Berenger}},\ }\href@noop {} {\bibfield  {journal} {\bibinfo  {journal} {J.
  Comput. Phys.}\ }\textbf {\bibinfo {volume} {114}},\ \bibinfo {pages} {185}
  (\bibinfo {year} {1994})}\BibitemShut {NoStop}%
\bibitem [{\citenamefont {Vial}\ \emph {et~al.}(2012)\citenamefont {Vial},
  \citenamefont {Zolla}, \citenamefont {Nicolet}, \citenamefont
  {Commandr\'{e}},\ and\ \citenamefont {Tisserand}}]{VialOE2012}%
  \BibitemOpen
  \bibfield  {author} {\bibinfo {author} {\bibfnamefont {B.}~\bibnamefont
  {Vial}}, \bibinfo {author} {\bibfnamefont {F.}~\bibnamefont {Zolla}},
  \bibinfo {author} {\bibfnamefont {A.}~\bibnamefont {Nicolet}}, \bibinfo
  {author} {\bibfnamefont {M.}~\bibnamefont {Commandr\'{e}}}, \ and\ \bibinfo
  {author} {\bibfnamefont {S.}~\bibnamefont {Tisserand}},\ }\href {\doibase
  10.1364/OE.20.028094} {\bibfield  {journal} {\bibinfo  {journal} {Opt.
  Express}\ }\textbf {\bibinfo {volume} {20}},\ \bibinfo {pages} {28094}
  (\bibinfo {year} {2012})}\BibitemShut {NoStop}%
\bibitem [{\citenamefont {Vial}\ \emph
  {et~al.}(2014{\natexlab{a}})\citenamefont {Vial}, \citenamefont {Zolla},
  \citenamefont {Nicolet},\ and\ \citenamefont {Commandr\'e}}]{VialPRA2014}%
  \BibitemOpen
  \bibfield  {author} {\bibinfo {author} {\bibfnamefont {B.}~\bibnamefont
  {Vial}}, \bibinfo {author} {\bibfnamefont {F.}~\bibnamefont {Zolla}},
  \bibinfo {author} {\bibfnamefont {A.}~\bibnamefont {Nicolet}}, \ and\
  \bibinfo {author} {\bibfnamefont {M.}~\bibnamefont {Commandr\'e}},\ }\href
  {\doibase 10.1103/PhysRevA.89.023829} {\bibfield  {journal} {\bibinfo
  {journal} {Phys. Rev. A}\ }\textbf {\bibinfo {volume} {89}},\ \bibinfo
  {pages} {023829} (\bibinfo {year} {2014}{\natexlab{a}})}\BibitemShut
  {NoStop}%
\bibitem [{\citenamefont {Vial}\ \emph
  {et~al.}(2014{\natexlab{b}})\citenamefont {Vial}, \citenamefont {Dem\'{e}sy},
  \citenamefont {Zolla}, \citenamefont {Nicolet}, \citenamefont {Commandr\'e},
  \citenamefont {Hecquet}, \citenamefont {Begou}, \citenamefont {Tisserand},
  \citenamefont {Gautier},\ and\ \citenamefont {Sauget}}]{VialMIM}%
  \BibitemOpen
  \bibfield  {author} {\bibinfo {author} {\bibfnamefont {B.}~\bibnamefont
  {Vial}}, \bibinfo {author} {\bibfnamefont {G.}~\bibnamefont {Dem\'{e}sy}},
  \bibinfo {author} {\bibfnamefont {F.}~\bibnamefont {Zolla}}, \bibinfo
  {author} {\bibfnamefont {A.}~\bibnamefont {Nicolet}}, \bibinfo {author}
  {\bibfnamefont {M.}~\bibnamefont {Commandr\'e}}, \bibinfo {author}
  {\bibfnamefont {C.}~\bibnamefont {Hecquet}}, \bibinfo {author} {\bibfnamefont
  {T.}~\bibnamefont {Begou}}, \bibinfo {author} {\bibfnamefont
  {S.}~\bibnamefont {Tisserand}}, \bibinfo {author} {\bibfnamefont
  {S.}~\bibnamefont {Gautier}}, \ and\ \bibinfo {author} {\bibfnamefont
  {V.}~\bibnamefont {Sauget}},\ }\href@noop {} {\bibfield  {journal} {\bibinfo
  {journal} {Submitted to JOSA B}\ } (\bibinfo {year}
  {2014}{\natexlab{b}})}\BibitemShut {NoStop}%
\end{thebibliography}
\end{document}